\documentclass[a4paper,aps,preprintnumbers,showpacs,twocolumn,amsmath,amssymb,prl]{revtex4}
\usepackage{mathrsfs}
\usepackage{amsmath}
\usepackage[dvips]{graphicx}
\usepackage{epsfig}
\usepackage[dvips]{graphics,graphicx}
\begin{document}

\title{Quantum phases of light coupled to a periodic Bose-Einstein condensate in a cavity}

\author{Aranya B.\ Bhattacherjee$^{1}$ and ManMohan$^{2}$}

\address{$^{1}$Department of Physics, ARSD College, University of Delhi (South Campus), New Delhi-110021, India}

\address{$^{2}$Department of Physics and Astrophysics, University of Delhi, Delhi-110007, India}

\begin{abstract}
We study a composite photon-Bose condensed atoms (BEC) system in an optical lattice confined in a cavity. We show that the system is able to produce an effective photon-photon repulsion in the presence of an external pump. We predict for the first time zero-temperature quantum phases of light coupled to BEC in the photon blockade regime in the presence of an external pump.
\end{abstract}


\maketitle

The search for interesting and potentially useful quantum-mechanical phenomena on a mesoscopic scale in condensed matter and atomic physics is a challenging task. The combination of cold atoms and cavity QED (quantum electrodynamics) is a conceptually new regime of cavity QED, in which all atoms occupy a single mode of a matter-wave field and couple identically to the cavity light field, sharing a single excitation. Experimental implementation of such systems has made significant progress \cite{Nagorny03,Sauer04,Anton05}. The atomic back action on the field introduces atom-field entanglement which modifies the associated quantum phase transition \cite{Maschler05}. Scattering of light from different atomic quantum states creates different quantum states of the scattered light, which can be distinguished by measuring the photon statistics of the transmitted light. \cite{Chen07,Mekhov07, Mekprl07,Mekpra07}. The band structure of the intracavity light field has been shown to influence the structural properties of the BEC \cite{Bhattacherjee}. Exotic phases of ultracold atoms in cavities have been predicted recently \cite{larson}.
The optical nonlinearities in the Jaynes-Cummings model generated due to the coupling between the atom and the photons leads to an effective photon-photon repulsion \cite{Birnbaum, Imam}. However it was shown that photon-photon repulsion degrades in the presence of many atoms \cite{Rebic}. Adding photons to a two-dimensional array of coupled optical cavities each containing a single two-level atom in the photon-blockade regime, a long-lived, strongly interacting dressed states of excitations (coupled atom-photons) are formed which can undergo at zero temperature a characteristics Mott insulator to superfluid quantum phase transition \cite{Greentree}.
Here we describe a coupled photon-atom system formed by adding photons to an one-dimensional array of Bose condensed atoms (BEC) confined in a cavity in the photon-blockade regime that exhibits photon-photon repulsion in the presence of an external pump. We search for a Hubbard-model type interactions in the coupled photon-atom system and predict the existence of quantum phases of composite photon-atom states at zero temperature.
\\
We consider an elongated cigar shaped Bose-Einstein condensate of $N$ two-level $^{87} Rb$ atoms in the $|F=1>$ state with mass $m$ and frequency $\omega_{a}$ of the $|F=1>\rightarrow |F'=2>$ transition of the $D_{2}$ line of $^{87} Rb$, strongly interacting with a quantized single standing wave cavity mode of frequency $\omega_{c}$. The internal cavity field is linked with the input and the output can then be calculated using the boundary condition at the cavity mirror. In order to create an elongated BEC, the frequency of the harmonic trap along the transverse direction should be much larger than one in the axial (along the direction of the optical lattice) direction. The system is also coherently driven by a laser field with frequency $\omega_{p}$ through the cavity mirror with amplitude $\eta$. The two sided cavity has two partially transparent mirrors with associated loss coefficient $\gamma_{1}$ and $\gamma_{2}$. The cavity decay is assumed to dominate over the spontaneous decay of all atoms thus allowing us to omit the effect of atomic decay. This system is modeled by a Bose-Hubbard (BH) type Hamiltonian in a rotating wave and dipole approximation\cite{Maschler05}. Adiabatically eliminating the excited state and retaining only the lowest band with nearest neighbor interaction,we obtain:

\begin{eqnarray}
&H& = \sum_{j}E_{0}\hat{b}_{j}^{\dagger}\hat{b}_{j}+E \left(\hat{b}_{j+1}^{\dagger}\hat{b}_{j}+\hat{b}_{j+1}\hat{b}_{j}^{\dagger} \right)\nonumber \\&+&\hbar U_{0}(\hat{a}_{j}^{\dagger}\hat{a}_{j}+1)\left\lbrace J_{0}\hat{b}_{j}^{\dagger}\hat{b}_{j}+J \left(\hat{b}_{j+1}^{\dagger}\hat{b}_{j}+\hat{b}_{j+1}\hat{b}_{j}^{\dagger} \right)\right\rbrace \nonumber \\&-&\hbar \Delta_{c} \hat{a}_{j}^{\dagger}\hat{a}_{j}-i\hbar \eta (\hat{a}_{j}-\hat{a}_{j}^{\dagger})+\dfrac{U}{2}\hat{b}_{j}^{\dagger}\hat{b}_{j}^{\dagger}\hat{b}_{j}\hat{b}_{j}.
\end{eqnarray}

\begin{figure}[t]
\hspace{-0.8cm}
\includegraphics[scale=0.7]{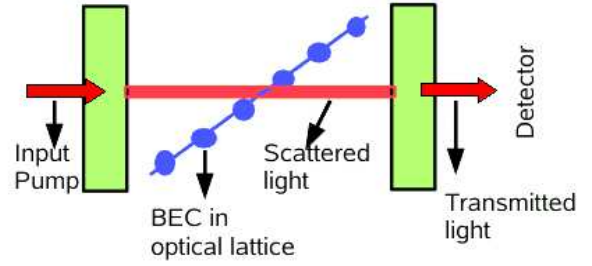}
\caption{Setup of the system. Atoms in a lattice are illuminated by the pump and the scattered light is transmitted through one of the mirrors and measured by a detector.}
\label{1}
\end{figure}

Here $U=\frac{4\pi a_{s}\hbar^{2}}{m}\int d^3 x  |w(\vec{r})|^{4}, E_{0}=\int d^3 x w(\vec{r}-\vec{r}_{j})( -\frac{\hbar^2 \nabla^{2}}{2m}) w(\vec{r}-\vec{r}_{j}),  E =\int d^3 x  w(\vec{r}-\vec{r}_{j})( -\frac{\hbar^2 \nabla^{2}}{2m}) w(\vec{r}-\vec{r}_{j \pm 1}),  J_{0}=\int d^3 x w(\vec{r}-\vec{r}_{j}) \cos^2(kx)w(\vec{r}-\vec{r}_{j}),  J =\int d^3 x w(\vec{r}-\vec{r}_{j}) \cos^2(kx) w(\vec{r}-\vec{r}_{j \pm 1})$ and $w(\vec{r}-\vec{r}_{j})$ is the Wannier function. The BH Hamiltonian of Eqn. (1) is different from the one derived in ref. \cite{Maschler05}, where the cavity photon operator is not dependent on the lattice site. A recent experiment \cite{Colombe} demonstrating that BEC atoms could be loaded into a single site of an intracavity optical lattice and deterministic transfer of the BEC atoms into successively addressed single sites of the lattice, each of which is differently coupled to the cavity photons provides a justification for the site dependence of the cavity photon operator. A single photon can be pumped into a well of the optical lattice by a low intensity pump laser with width less than the lattice spacing so that the spatial overlap of the pump and the lattice is no larger than the lattice spacing. The first photon will form a composite photon-atom system and a secong photon that comes in will experience a repulsion unless there is a second atom to form a second photon-atom system. In ref. \cite{Colombe}, a probe laser with $3.9 \mu m$ width and lattice spacing of $6.4 \mu m$ was utilized. The parameter $U_{0}=\dfrac{g_{0}^{2}}{\Delta_{a}}$ is the optical lattice barrier height per photon and represents the atomic backaction on the field \cite{Maschler05}. Here we will always take $U_{0}>0$. In this case the condensate is attracted to the nodes of the light field and hence the lowest bound state is localized at these positions. Along $x$, the cavity field forms an optical lattice potential of period $\lambda/2$ and depth $\hbar U_{0}(\hat{a}_{j}^{\dagger}\hat{a}_{j})$ at each lattice site. Here $\Delta_{a}=\omega_{p}-\omega_{a}$ and $\Delta_{c}=\omega_{p}-\omega_{c}$ are the large atom-pump and cavity-pump detuning, respectively and  $\Delta_{c}>\Delta_{a}$. In this work we will consider only the case  $\Delta_{a}>0$. The atom-field coupling is written as $g(x)=g_{0} \cos(kx)$. Here $\hat{a}_{j}$ and $\hat{b}_{j}$ are the annihilation operator for a cavity photon and the bosonic atom in the $j^{th}$ site of the optical lattice respectively. Here $a_{s}$ is the two body $s$-wave scattering length. The onsite energies $J_{0}$ and $E_{0}$ are set to zero. The Heisenberg equation of motion for the bosonic field operator $\hat{b}$ is

\begin{eqnarray}
\dot{\hat{b}}_{j}&=&-iU_{0}\left( 1+\hat{a}_{j}^{\dagger} \hat{a}_{j} \right)J\left\lbrace \hat{b}_{j+1}+\hat{b}_{j-1} \right\rbrace-\dfrac{iE}{\hbar}\left\lbrace\hat{b}_{j+1}+\hat{b}_{j-1}  \right\rbrace\nonumber \\&-& \dfrac{iUn_{0}}{\hbar}\hat{b}_{j}\;
\end{eqnarray}

The behaviour of the internal cavity mode is obtained from the quantum-Langevin equation which for a single-mode cavity becomes

\begin{eqnarray}
\dot{\hat{a}}_{j}&=&-iU_{0}\left\lbrace J_{0}\hat{b}_{j}^{\dagger}\hat{b}_{j}+J  \left(\hat{b}_{j+1}^{\dagger}\hat{b}_{j}+\hat{b}_{j+1}\hat{b}_{j}^{\dagger} \right)\right\rbrace \hat{a}_{j}+\eta\nonumber \\&+&i\Delta_{c} \hat{a}_{j}-\dfrac{\gamma_{1}}{2} \hat{a}_{j}-\dfrac{\gamma_{2}}{2} \hat{a}_{j}+\sqrt{\gamma_{1}}\hat{a}_{in}+\sqrt{\gamma_{2}}\hat{b}_{in}\;
\end{eqnarray}

Here $\hat{a}_{in}$ and $\hat{b}_{in}$ are the external input fields incident from the two mirrors.. Equation (3) and (4) represents a set of coupled equations describing the dynamics of the compound system formed by the condensate and the optical cavity. We will work in the bad cavity limit, where typically, $\gamma_{1}$ and $\gamma_{2}$ are the fastest time scale ( this means that the cavity decay rates are much larger than the oscillation frequency of bound atoms in the optical lattice of the cavity ). In this limit the intracavity field adiabatically follows the condensate wavefunction, and hence we can put $\dot{\hat{a}}_{j}=0$. We treat the BEC within the mean field framework and assume the tight binding approximation where we replace $\hat{b}_{j}$ by $\phi_{j}$ and look for solutions in the form of Bloch waves $\phi_{j}=u_{k}exp(ikjd)exp(-i\mu t/ \hbar)$. Here $\mu$ is the chemical potential, $d$ is the periodicity of the lattice and $\dfrac{1}{I}\sum_{j}\hat{b}_{j}^{\dagger} \hat{b}_{j}=|u_{k}|^{2}=n_{0}$ (atomic number density), $I$ is the total number of lattice sites and $\sum_{j} n_{0}=N$ (total number of atoms). The relationship between the input and output modes may be found from using the boundary conditions at each mirror,
$\tilde{a}_{out}(\omega)+\tilde{a}_{in}(\omega)=\sqrt{\gamma_{1}} \tilde{a}_{j}(\omega)$ and $\tilde{b}_{out}(\omega)+\tilde{b}_{in}(\omega)=\sqrt{\gamma_{2}} \tilde{a}_{j}(\omega)$ \cite{Collett}. In frequency space, we obtain

\begin{equation}
\tilde{a}_{out}(\omega)=\dfrac{\sqrt{\gamma_{1}}\eta+(\gamma_{1}/2-\gamma_{2}/2+i \Delta)\tilde{a}_{in}(\omega)+ \sqrt{\gamma_{1} \gamma_{2}}\tilde{b}_{in}(\omega)}{\left\lbrace \gamma_{1}/2+\gamma_{2}/2-i \Delta^{'} \right\rbrace }.
\end{equation}

Where $\tilde{a}_{out(in)}(\omega)=\dfrac{1}{\sqrt{2 \pi}} \int_{-\infty}^{\infty} e^{i \omega t} \hat{a}_{out(in)}(t) dt$ and $\Delta^{'}= \Delta_{c}+\omega-2Jn_{0}U_{0}\cos(kd)$.We find that due to the atomic backaction, the quantum state of the output cavity field develops a band structure due to the strong coupling with the condensate, analogous to photonic band gap materials \cite{Deutsch94}. The average photon number $<\hat a^{\dagger}_{out} \hat a_{out}>$ measures the light transmission spectra and is different for the Mott insulator (MI) and the superfluid phase (SF) \cite{Mekhov07}. If the two mirrors are the same, $\gamma_{1}=\gamma_{2}=\gamma$ and near resonance $\Delta^{'} \approx 0$, we find: $\tilde{a}_{out}(\omega) \approx \dfrac{\sqrt{\gamma}\eta+ \gamma \tilde{b}_{in}(\omega)}{\left\lbrace \gamma-i \Delta^{'} \right\rbrace }$.
The resonance point is one where the superfluid fraction is minimum \cite{Bhattacherjee}. This shows that the cavity now behaves like a shifted through-pass Lorentzian filter. The input field will be completely reflected if $\Delta^{'}>> \gamma$ (when the atoms are in the deep superfluid regime), $\tilde{a}_{out}(\omega)\approx - \tilde{a}_{in}(\omega)$. The state of the BEC together with the cavity parameters control the output field of the cavity.

The Hamiltonian of Eqn.(1) in the mean field of the atoms is written as

\begin{eqnarray}
H^{MF}&=& \sum_{j}  2En_{0} \cos{kd}+2\hbar U_{0} J n_{0} \cos{kd}(1+\hat{a}_{j}^{\dagger} \hat{a}_{j})\nonumber
\\&-&\hbar \Delta_{c} \hat{a}_{j}^{\dagger} \hat{a}_{j}-i \hbar \eta (\hat{a}_{j}-\hat{a}_{j}^{\dagger})+\dfrac{U }{2}|u_{k}|^{4}.
\end{eqnarray}

\begin{figure}[t]
\hspace{-0.8cm}
\includegraphics[scale=0.8]{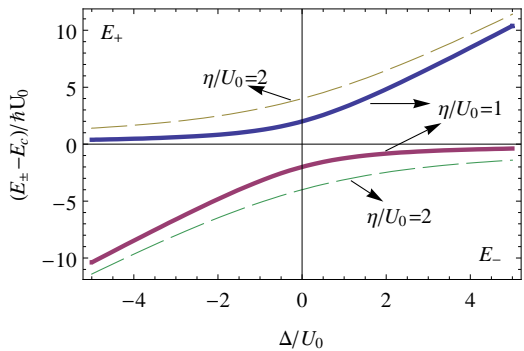}
\caption{Dimensionless eigenspectrum, $(E_{\pm}-E_{c})/\hbar U_{0}$ as a function of $\Delta/U_{0}$ for $n=1$. The eigenspectrum splits into two branches , corresponding to the dressed states, $E_{+,n}$ (upper branch) and $E_{-,n}$ (lower branch).The two branches anti-cross at $\Delta/U_{0}=0$, with the splitting increasing with increasing pump strength.}
\label{2}
\end{figure}

\begin{figure}[t]
\hspace{-0.5cm}
\includegraphics[scale=0.8]{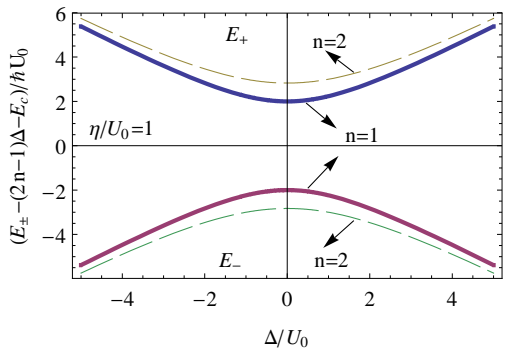}
\caption{Dimensionless eigenspectrum, $(E_{\pm}-E_{c}-(2n-1)\Delta)/\hbar U_{0}$ versus $\Delta/U_{0}$ for a fixed pump strength and $n=1$(solid lines) and $n=2$(dashed lines). The two branches anti-cross at $\Delta/U_{0}=0$, with the splitting increasing with increasing excitation number, $n$.}
\label{3}
\end{figure}

Analogous to the Jaynes-Cummings model, the mean field Hamiltonian $H^{MF}$ leads to photon-photon repulsion in the presence of the pump. The conserved particles in out model are not the photons but composite photon-atom states. The eigenvalues of the $H_{j}^{MF}$ (mean-field Hamiltonian at each lattice site)in the basis $|g,n>$ and $|g,n-1>$ is found as,

\begin{equation}
E_{\pm,n}=\dfrac{(2n-1)\hbar \Delta}{2} \pm \dfrac{1}{2}\sqrt{\hbar^{2}\Delta^{2}+4n\hbar^{2} \eta^{2}}+\dfrac{E_{c}}{2}
\end{equation}

where $|g,n>$ is the ground state of the composite atom-photon system with $n$ excitations. Here $\Delta=2U_{0}n_{0}J \cos{kd}-\Delta_{c}$ and $E_{c}=2n_{0}(E \cos{kd}+\hbar U_{0}J \cos{kd})+U/2|u_{k}|^{2}$. In Fig.2, we show the dimensionless eigenenergies $(E_{\pm}-E_{c})/\hbar U_{0}$ as a function of $\Delta/U_{0}$ for $n=1$. The bold lines are for $\eta/U_{0}=1$ and the dashed lines are for $\eta/U_{0}=2$. The increasing energy separation with increasing pump strength is an indication of photon-photon repulsion. In Fig.3, a plot of $(E_{\pm}-E_{c}-(2n-1)\Delta)/\hbar U_{0}$ versus $\Delta/U_{0}$ for a fixed pump strength and $n=1$(solid lines) and $n=2$(dashed lines). The on-site photonic repulsion is evinced by the increasing energy separation with $n$.
In the absence of the pump, the photon-photon replusion is absent.

The effective Hamiltonian for our extended Hubbard-like system is given by a combination of the mean-field Hamiltonian with photon hopping between wells and the chemical potential term. We introduce a superfluid order parameter $\psi=<\hat{a}_{j}>$, which we take to be real and use the decoupling approximation, $\hat{a}_{j}^{\dagger}\hat{a}_{j}=<\hat{a}_{j}^{\dagger}>\hat{a}_{j}+<\hat{a}_{j}>\hat{a}_{j}^{\dagger}-<\hat{a}_{j}^{\dagger}> <\hat{a}_{j}>$. The resulting effective mean-field Hamiltonian can be written as a sum over single sites

\begin{eqnarray}
H_{eff}^{MF}&=&\sum_{j}(H_{j}^{MF}-z t \psi (\hat{a}_{j}^{\dagger}+\hat{a}_{j})\nonumber \\&+& z t \psi^{2}-\mu(\hat{a}_{j}^{\dagger}\hat{a}_{j}+n_{0})),
\end{eqnarray}

where $t$ is the nearest neighbor hopping energy of the photons coupled with the atoms (actually it is the excitations which hop from one well to the other) and $\mu_{j}$ is the chemical potential at site $j$. We assume zero disorder such that $\mu_{j}=\mu$ for all sites. Here $z=2$ is the number of nearest neighbors. To obtain the system's zero-temperature phase diagram, we use the procedure of ref. \cite{Oosten}. This Hamiltonian is diagonal with respect to the site index $j$, so we can use an effective onsite Hamiltonian. If we introduce $\bar{U}_{0}=U_{0}/zt$, $\bar{\mu}=\mu/zt$, $\bar{H}_{j}^{MF}=H_{j}^{MF}/zt$, we find $H_{eff}^{MF}=H^{0}+\psi V$ with $H^{0}=\bar{H}_{j}^{MF}+\psi^{2}-\bar{\mu}(\hat{a}_{j}^{\dagger}\hat{a}_{j}+n_{0})$ and $V=-(\hat{a}_{j}^{\dagger}+\hat{a}_{j})$.
The unperturbed ground state energy of the state with exactly $n$ particles is $E_{g,n}^{0}=\bar{E}_{-,n}-\bar{\mu}n$, where $\bar{E}_{-,n}=E_{-,n}/zt$. We only need to consider the negative branch for the purpose of determining the ground state since $E_{-,n}<E_{+,n}$. A change in the total number of excitations per site will occur when $\bar{E}_{-,n+1}-\bar{\mu}(n+1)=\bar{E}_{-,n}-\bar{\mu}n$. We can determine the critical chemical potential, $\bar{\mu}_{c}(n)$, where the system will change from $n$ to $n+1$ excitations per site as
$\bar{\mu}_{c}(n)=\bar{\Delta}+(\bar{\chi}_{n}-\bar{\chi}_{n+1})$,where $\bar{\chi}_{n}=\sqrt{\Delta^{2}/4+n \hbar^{2} \eta^{2}}/zt$.

\begin{figure}[t]
\hspace{-0.5cm}
\includegraphics[scale=0.8]{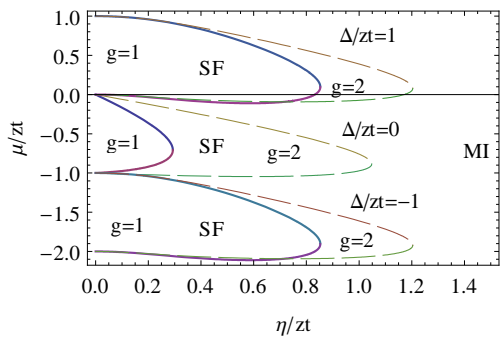}
\caption{Phase diagram of the coupled periodic photon-BEC system in the cavity in the presence of the pump. The vertical axis shows $\mu/zt$ and the horizontal axis shows $\bar{\eta}$. The plot is shown for $g=1,2$ and $\bar{\Delta}=-1,0,1$. Dominating the right-hand edge (where photonic repulsion dominates at higher pump strength) is the Mott-insulator (denoted by MI) and the superfluid phase is found on the left-hand edge (denoted by SF).}
\label{4}
\end{figure}

Since the interaction $V$ couples only to states with one more or less excitations than in the ground state, we find the second order correction to energy

\begin{equation}
E_{g}^{(2)}=\dfrac{g}{\bar{\Delta}+(\bar{\chi}_{g}-\bar{\chi}_{g-1})-\bar{\mu}}+\dfrac{g+1}{-\bar{\Delta}+(\bar{\chi}_{g+1}-\bar{\chi}_{g})+\bar{\mu}}
\end{equation}

According to the Landau procedure for second-order phase transitions, we write the ground state as an expansion in $\psi$ as $E_{g}(\psi)=a_{0}(g,\bar{\mu},\bar{\eta},\bar{\Delta})+a_{2}(g,\bar{\mu},\bar{\eta},\bar{\Delta})\psi^{2}+O(\psi^{4})$.
$E_{g}(\psi)$ is minimized as a function of the superfluid order parameter $\psi$. We find that $\psi=0$ when $a_{2}(g,\bar{\mu},\bar{\eta},\bar{\Delta})>0$ and that $\psi \neq 0$ when $a_{2}(g,\bar{\mu},\bar{\eta},\bar{\Delta})<0$. This means that $a_{2}(g,\bar{\mu},\bar{\eta},\bar{\Delta})=0$ signifies the boundary between the superfluid and insulator phases of light. This yields

\begin{equation}
\bar{\mu}_{\pm}=\dfrac{1}{2}(2 \bar{\Delta}-\bar{\chi}_{g+1}+\bar{\chi}_{g-1}-1)\pm \dfrac{1}{2}\sqrt{1+\Sigma^{2}+2(2n+1)\Sigma},
\end{equation}

where $\Sigma=\bar{\chi}_{g+1}+\bar{\chi}_{g-1}-2\bar{\chi}_{g}$. The subscript $\pm$ denotes the upper and lower halves of the superfluid regions of phase space. Fig.3 shows a plot of Eqn.(27) for $g=1,2$ and $\bar{\Delta}=-1,0,1$ as a function of $\bar{\eta}$. By equating $\bar{\mu}_{-}$ and $\bar{\mu}_{+}$ we can find the point of largest $\bar{\eta}_{max}$ for each superfluid region (SF). The dynamics illustrated in Fig.3 is extremely rich. Dominating the right-hand edge (where photonic repulsion dominates at higher pump strength) is the Mott-insulator (denoted by MI) and the superfluid phase is found on the left-hand edge (denoted by SF).The size of the superfluid region is found to increase with the excitations. On increasing the pump strength, the optical potential increases and hence the atoms loose their superfluid behavior \cite{Bhattacherjee}. The transition from the $SF$ to the $MI$ phase of the excitations occurs at $\eta_{max}$. In ref. \cite{Bhattacherjee} it was shown that the superfluid fraction of the cold atoms shows a minimum at $\Delta=0$. In accordance with this result, we find that the size of the superfluid lobes of the composite photon-atom system is minimum at $\bar{\Delta}=0$.
To find the eigenenergies and to experimentally identify the various phases of the coupled BEC-cavity system, one can perform a transmission spectroscopy with the scattered light by direct read out of the number of photons coming out of the cavity through Eqn.(6). The transmission of the scattered light is monitored as a function of $\Delta$. To probe the system in the weak excitation limit, the mean intracavity photon number should be below $\gamma^{2}/2g_{0}^{2} \approx 0.04$ \cite{Ferd07}. Photon loss can be minimized by using high-Q cavities and thus ensuring that the light field remains quantum-mechanical for the duration of the experiment. It is important that the characteristic time-scales of coherent dynamics are significantly faster than those associated with losses (the decay rate ($\gamma$) of state-of-art optical cavities is typically 17 kHz \cite{Klinner06}).
In summary, we have shown that a system of periodic Bose condensed atoms coupled to cavity photons in the photon blockade regime exhibits a rich strongly correlated dynamics. The composite photon-atom states thus formed undergoes a characteristic Mott insulator to superfluid quantum phase transition. Because of the Mott phase's robustness, the present system provides a useful platform for developing concepts in quantum information processing.

\end{document}